\documentclass{ws-ijtaf}
\usepackage{psfig}

\begin{document}


%
%

\title{The Wick theorem for non-Gaussian distributions
       and its application for noise filtering 
       of correlated $q$-Exponentialy distributed random variables}

\author{Przemys\mbox{\l}aw Repetowicz and Peter Richmond}

\address{Department of Physics, Trinity College Dublin 2, Ireland}

\maketitle


\begin{abstract}
We derive the Wick theorem for the $q$-Exponential distribution.
We use the theorem to derive a numerical algorithm for
finding parameters of the correlation matrix of q-Exponentialy
distributed random variables given empirical spectral moments of the time series.
\end{abstract}

\keywords{Many-point correlation functions; Wick theorem; Portfolio management;} 

\section{Introduction}
$q$-Exponential distributions ($q$-Exponentials defined in the next section)
are used\cite{KleinertBook} in modeling distributions of stocks.
The $q$-Exponentials possess two desirable features namely 
exhibit power law tails in the high end of the distribution and tend toward a Gaussian 
when $q = 1 + 1/K \rightarrow 1$. These facts provide enough motivation
to derive a variant of a Wick theorem for linear combinations
of independent identically distributed (iid) $q$-Exponentials 
(correlated $q$-Exponentials).

Recent growth of interest in applications of physics to economics
and to the theory of finance (econophysics) yields the Wick theorem useful
for the following purposes:
\begin{enumerate}
\item Deriving exact relations between spectra of eigenvalues of 
financial covariance matrices related to $q$-exponentially distributed time series
and of estimators of covariance matrices. This will be a generalization of 
existing exact relations \cite{JurkiewiczRes} for Gaussian distributed time series.

\item Deriving algorithms \cite{Sornette} for optimizing a portfolio 
(minimizing the variance and/or the higher moments of a portfolio
subject to a presumed return from a portfolio) of stocks whose time series are $q$-exponentially
distributed.

\item Verifying empirical findings \cite{StanleyEconophys} regarding a very fast time decay 
of the autocorrelation function in time series and regarding a power-law time dependence
of the autocorrelation function of the volatility; investigating many point correlations
in financial time series.
\end{enumerate}

\section{The Wick theorem}
The objective of this section is to derive the Fourier transform 
of $N$ correlated $q$-Exponentially distributed variables.

A random variable $X$ is $q$-Exponentially distributed,
($X \sim q-\mbox{Exp}$) if $q = 1 + 1/K$ 
and the probability density function (pdf) $D_K(x)$
of $X$ reads:
\begin{equation}
D_K(x) = \frac{N_K}{\sqrt{2 \pi \sigma_K^2}} e_K^{-x^2/(2 \sigma_K^2)}
\label{eq:Definition}
\end{equation}
where $e_K^z := (1 - z/K)^{-K}$, 
$N_k = \Gamma(K)/(\sqrt{K} \cdot \Gamma(K - 1/2))$
and $\sigma_K = \sigma \sqrt{(K - 3/2)/K}$.

The pdf (\ref{eq:Definition}) has (equation (20.52) in \cite{KleinertBook})
is a continuous superposition of Gaussians and has a following 
integral representation:
\begin{equation}
D_K(x) = \frac{1}{\mathfrak{D}^D}
\int_{\mathbb{R}^D} d^D \xi e^{-\xi^2 (D+1)/2} \cdot 
\frac{e^{-\xi^2 x^2/(2 \sigma_D^2)}}{\sqrt{2\pi \sigma_D^2 / \xi^2}}
\label{eq:Projection}
\end{equation}
where $\mathfrak{D} := \sqrt{2\pi/(D+1)}$ and 
$\sigma_D = \sigma \sqrt{(D-2)/(D+1)} = \sigma_K$ and the integral runs over 
the $D$-dimensional space $(\xi_1,\dots,\xi_D)$ with:
\begin{equation}
D = 2 K - 1 
\end{equation}

We define correlated $q$-Exponentials as linear combinations of iid $q$-Exponentials. 
Therefore the joint pdf $\rho_{\vec{X}}(\vec{x})$ of $N T$ correlated 
$q$-Exponentials
\begin{equation}
X_{i,t} = \sum_{j=1}^N\sum_{t=1}^T O_{i,t}^{j,\theta} Y_{j,\theta}
\end{equation}
where $i=1,\ldots,N$, $t=1,\ldots,T$ and $Y_{j,\theta} \sim q-\mbox{Exp}$ are iid,
reads:
\begin{eqnarray}
\rho_{\vec{X}}(\vec{x}) &=& 
 \int_{\mathbb{R}^{N T}}
\delta(\vec{x} - \underline{\underline{O}} \vec{y}) 
\prod_{i=1}^N \prod_{t=1}^T D_K(y_{i,t}) d y_{i,t} \\
&=& (\mbox{det}\underline{\underline{O}})^{-1}
\mathop{\prod_{i=1,\dots,N}}_{t=1,\dots,T}
D_K(\mathop{\sum_{j=1,\dots,N}}_{\theta=1,\dots,T} (O^{-1})_{i,t}^{j,\theta} x_{j,\theta})
\end{eqnarray}
Here we call 
$\underline{\underline{O}} := 
\mathop{\left\{ O_{i,t}^{j,\theta} \right\}}
_{\begin{array}{l} i=1,\dots,N \\ t=1,\dots,T \end{array}}$
a rotation tensor.
We take $\vec{k} := \left\{k_{i,t}\right\}$ and we calculate the Fourier transform 
$\kappa(\vec{k}) = \mathcal{F}_{\vec{X}}\left[\rho_{\vec{X}}\right](\vec{k})
:= \int_{\mathbb{R}^{N T}} d^{N T} \vec{x} \rho_{\vec{X}}(\vec{x}) e^{\imath \vec{k} \vec{x}}$. 
It reads:

\begin{eqnarray}
\lefteqn{
(\mathfrak{D}^{D N T} \mbox{det}\underline{\underline{O}}) \kappa(\vec{k}) =
(\mathfrak{D}^{D N T} \mbox{det}\underline{\underline{O}}) 
\int_{\mathbb{R}^{N T}} d^{N T} \vec{x} \rho_{\vec{X}}(\vec{x}) e^{\imath \vec{k} \vec{x}} = 
 } \label{eq:FourierTransformStart}\\
&& 
\!\!\!\!\!\!\!\!\!\!\!\!\!\!\!\!\!\!\!\!\!\!
\int_{\mathbb{R}^D} \dots \int_{\mathbb{R}^D} \mathop{\prod_{i=1,\dots,N}}_{t=1,\dots,T} d\xi_{i,j}
\left(
\mathop{\prod_{i=1,\dots,N}}_{t=1,\dots,T} \frac{e^{-\xi_{i,t}^2 (D+1)/2}}
                                           {\sqrt{2\pi \sigma_D^2/\xi_{i,t}^2}}
\right)
\int_{\mathbb{R}^{N T}} d^{N T}\vec{x} 
\exp\left[
\left(-\frac{\xi_{i,t}^2}{2 \sigma_D^2}\right) 
\left(\mathop{\sum_{j=1,\dots,N}}_{\theta=1,\dots,T} (O^{-1})_{i,t}^{j,\theta} x_{j,\theta}\right)^2
\right]
	      e^{\vec{\imath k}\cdot\vec{x}}  \nonumber \\
&&
\!\!\!\!\!\!\!\!\!\!\!\!\!\!\!\!\!\!\!\!\!\!
\int_{\mathbb{R}^D} \dots \int_{\mathbb{R}^D} \mathop{\prod_{i=1,\dots,N}}_{t=1,\dots,T} d\xi_{i,j}
\left(
\mathop{\prod_{i=1,\dots,N}}_{t=1,\dots,T} 
\frac{e^{-\xi_{i,t}^2 (D+1)/2}}{\sqrt{2\pi \sigma_D^2/\xi_{i,t}^2}}
\right)
\int_{\mathbb{R}^{N T}} d^{N T}x 
\exp\left\{ 
 - \frac{1}{2} \vec{x}^T \cdot \underline{\underline{C}}^{-1}(\vec{\xi}) \cdot \vec{x} 
 + \imath \vec{k} \vec{x} 
 \right\}
\label{eq:FourierTransformEnd}
\end{eqnarray}
where we introduced $\underline{\underline{C}}(\vec{\xi}) := 
\underline{\underline{O}} \cdot \underline{\underline{D}} \cdot \underline{\underline{O}}^T$
(a correlation tensor)
that is related to the rotation tensor $\underline{\underline{O}}$
and to a diagonal tensor 
$D_{i,t}^{j,\theta} = \delta_{i,j}\delta_{t,\theta} \sigma_D^2/\xi_{i,t}^2$.
The transposition $^T$ operation is defined as 
$(O^T)_{i,t}^{j,\theta} := O_{j,\theta}^{i,t}$.
This means that:
\begin{equation}
C_{i,t}^{j,\theta}(\vec{\xi}) := \mathop{\sum_{p=1,\dots,N}}_{\lambda=1,\dots,T} 
                                 O_{i,t}^{p,\lambda} \frac{\sigma_D^2}{\xi_{p,\lambda}^2}
				 O_{p,\lambda}^{j,\theta} 
\label{eq:DefinitionCorrelVar}
\end{equation}

The last integral on the right hand side in (\ref{eq:FourierTransformEnd})
is evaluated by ``completing to a square'' and it reads:
\begin{eqnarray}
\lefteqn{
(2 \pi)^{N T/2} \left(\mbox{det}(\underline{\underline{C}}^{-1}(\vec{\xi}))^{-1/2}\right) \cdot 
\exp\left\{ -\frac{1}{2} \vec{k}^T \underline{\underline{C}}(\vec{\xi}) \vec{k} \right\}
} \\
&&= (2 \pi)^{N T/2} \left(\mbox{det}\underline{\underline{O}}\right)
\sigma_D^{N T} \left( \prod_{j=1}^N\prod_{\theta=1}^T \xi_{j,\theta}^{-1} \right) \cdot
\exp\left\{ -\frac{1}{2} \vec{k}^T \underline{\underline{C}}(\vec{\xi}) \vec{k} \right\}
\end{eqnarray}
Therefore the Fourier transform (\ref{eq:FourierTransformStart}) reads:

\begin{eqnarray}
\lefteqn{
\kappa(\vec{k}) 
:= } \\
&&
\frac{(2 \pi)^{N T/2} \sigma_D^{N T}}{\mathfrak{D}^{D N T}}
\int_{\mathbb{R}^D} \dots \int_{\mathbb{R}^D} 
\mathop{\prod_{j=1,\dots,N}}_{\theta=1,\dots,T} d\xi_{j,\theta}
\left( \frac{ e^{-\xi_{j,\theta}^2 (D+1)/2} }{\sqrt{2\pi \sigma_D^2/\xi_{j,\theta}^2}} 
(\xi_{j,\theta})^{-1}
\right)
\cdot
\exp\left\{ -\frac{1}{2} \vec{k}^T \underline{\underline{C}}(\vec{\xi}) \vec{k} \right\} 
\nonumber \\
&&= \frac{1}{\mathfrak{D}^{D N T}}
\int_{\mathbb{R}^D} \dots \int_{\mathbb{R}^D} 
\mathop{\prod_{j=1,\dots,N}}_{\theta=1,\dots,T} d\xi_{j,\theta}
\left(
e^{-\xi_{j,\theta}^2 (D+1)/2}
\right)
\exp\left\{ -\frac{1}{2} \vec{k}^T \underline{\underline{C}}(\vec{\xi}) \vec{k} \right\} \\
&&= \left< \exp\left\{ -\frac{1}{2} \vec{k}^T \underline{\underline{C}}(\vec{\xi}) \vec{k} \right\}
\right>_{\vec{\xi}}
\label{eq:FourierTransformFinal}
\end{eqnarray}
The weight $\omega(\vec{\xi})$ used in the average $\left<\right>_{\vec{\xi}}$
in (\ref{eq:FourierTransformFinal}) reads:
\begin{equation}
\omega(\vec{\xi}) = 
\left(\mathcal{N}\right)^{N T}
\mathop{\prod_{j=1,\dots,N}}_{\theta=1,\dots,T} \xi_{j,\theta}^{D-1} 
\exp\left\{ -\frac{\xi_{j,\theta}^2 (D+1)}{2}\right\}
\label{eq:Weight}
\end{equation}
where $\mathcal{N} := ((D+1)^{D/2})/(2^{D/2 - 1}\Gamma(D/2))$.

In (\ref{eq:Weight}) we expressed the integral over $\xi_{j,\theta}$
in radial coordinates
$\int_{\mathbb{R}^D} d^D\xi = 2(\pi)^{D/2}/(\Gamma(D/2)) \int_0^\infty \xi^{D-1} d\xi$.

Now we differentiate (\ref{eq:FourierTransformFinal}) $2k$ times
with respect to variables $k_{j(1),\theta(1)},\dots,k_{j(2k),\theta(2k)}$
and evaluate the result at $\vec{k} = \vec{0}$. We get:

\begin{eqnarray}
\lefteqn{
\left<X_{j(1),\theta(1)} \cdot \dots \cdot X_{j(2k),\theta(2k)}\right> =
\frac{1}{\imath^{2k}}
\frac{\partial}{\partial k_{j(1),\theta(1)}} \cdot \dots \cdot
\frac{\partial}{\partial k_{j(2 k),\theta(2 k)}}
\left< \kappa(\vec{k}) 
\right> \left|_{\vec{k} = \vec{0}} \right.
} \\
&&=
\frac{1}{(-1)^{k}}
\frac{\partial}{\partial k_{j(1),\theta(1)}} \cdot \dots \cdot
\frac{\partial}{\partial k_{j(2 k),\theta(2 k)}}
\left<
\frac{1}{2^k k!}
\left(-\mathop{\sum_{i,j=1,\dots,N}}_{i,\theta=1,\dots,T} C_{i,t}^{j,\theta}(\vec{\xi}) k_{i,t} k_{j,\theta} \right)^k  \right>\left|_{\vec{k} = \vec{0}} \right. 
 \label{eq:ManyPointCorrel0} \\
&&= 
\sum\limits_{\sigma = \mathfrak{c}^{(1)}\circ \dots \circ \mathfrak{c}^{(k)}}
\left< 
\prod_{q=1}^k 
C_{j(\mathfrak{c}^{(q)}_1),\theta(\mathfrak{c}^{(q)}_1)}^{j(\mathfrak{c}^{(q)}_2),\theta(\mathfrak{c}^{(q)}_2)}(\vec{\xi})
\right>_{\vec{\xi} \in \mathbb{R}^{N T \cdot D}}
       \label{eq:ManyPointCorrel} \\
&&=\sum\limits_{\sigma = \mathfrak{c}^{(1)}\circ \dots \circ \mathfrak{c}^{(k)}}
\left< 
\prod_{q=1}^k \left[
              O_{j(\mathfrak{c}^{(q)}_1),\theta(\mathfrak{c}^{(q)}_1)}^{p(q),\lambda(q)}
              O_{p(q),\lambda(q)}^{j(\mathfrak{c}^{(q)}_2),\theta(\mathfrak{c}^{(q)}_2)}
	      \mathop{\prod_{l=1,\dots,N}}_{m=1,\dots,T}
	      (
	      \mathcal{N} \int_0^\infty \xi^{D - 2\nu(l,m) - 1} e^{-\xi^2 \frac{(D+1)}{2}} d\xi
	      )
	      \right]
	      \right> \label{eq:ManyPointCorrel1} \\
&&=\sum\limits_{\sigma = \mathfrak{c}^{(1)}\circ \dots \circ \mathfrak{c}^{(k)}}
\left< 
\prod_{q=1}^k \left[
              O_{j(\mathfrak{c}^{(q)}_1),\theta(\mathfrak{c}^{(q)}_1)}^{p(q),\lambda(q)}
              O_{p(q),\lambda(q)}^{j(\mathfrak{c}^{(q)}_2),\theta(\mathfrak{c}^{(q)}_2)}
	      \mathop{\prod_{l=1,\dots,N}}_{m=1,\dots,T}
	      (
	      \left(\frac{D+1}{2}\right)^{\nu(l,m)} \frac{\Gamma((D/2 - \nu(l,m))}{\Gamma(D/2)}
	      )
	      \right]
	      \right> \label{eq:ManyPointCorrel2} 
\end{eqnarray}
where the sum in (\ref{eq:ManyPointCorrel}) runs over all $2k$-permutations $\sigma$
that are composed exclusively of cycles 
$\mathfrak{c}^{(i)} := (\mathfrak{c}^{(i)}_1, \mathfrak{c}^{(i)}_2)$ 
for $i=1,\dots,k$ of lengths two.
Since the functions $C_{i,t}^{j,\theta}$ are symmetric with respect to exchanging
the above and lower pairs
and multiplication is commutative the whole of $(2k)!$ terms in the sum in (\ref{eq:ManyPointCorrel0})
decomposes into $(2 k)!/(2^k k!) = (2k - 1)!!$ distinct terms
who occur $(2^k k!)$ times each.
In this way the factor $2^k k!$ in the denominator in
 (\ref{eq:ManyPointCorrel0}) cancels out.
In (\ref{eq:ManyPointCorrel1}) we used the definition (\ref{eq:DefinitionCorrelVar})
of the correlation tensor and we introduced new indices $p(q)=1,\dots,N$ and $\lambda(q)=1,\dots,T$
Since the average over $\vec{\xi}$ in (\ref{eq:ManyPointCorrel})
consists in performing integrals of the kind
$\mathcal{N} \int_{\mathbb{R}^d} \prod_{q=1}^k \xi_{p(q),\lambda(q)}^{-2} \omega(\vec{\xi}) d\vec{\xi}$ it is readily seen that the result
(\ref{eq:ManyPointCorrel1}) is expressed via a number $\nu(l,m)$
that depends on the sequence $\left\{p(q),\lambda(q)\right\}$ and that 
is equal to the multiplicity of the pair $(l,m)$ in that sequence:
\begin{equation}
\nu(l,m) = \#\left\{ \mbox{pairs $(p_q,\lambda_q)$ such that $p_q=l$ and $\lambda_q=m$} \right\}
\end{equation}

\section{Covariances their estimators and noise filtering}
In this section we discuss 
definitions of averages over ensembles of stochastic variables
(resolvents) whose properties may be compared to measured properties 
of financial time series.


\begin{definition}
The resolvent function is a complex function $G(z)$ such that:
\begin{equation}
$\mbox{Im$\left[G(z + \imath \epsilon)\right] = \delta_\epsilon(z)$}$, 
\quad \epsilon > 0
\end{equation}
where $\delta_\epsilon(z)$ is a representation of the delta function.
\end{definition}

The function $G(z)$ is used for finding the density of eigenvalues $\rho_\Lambda(\lambda)$
of the covariance $\left< \mathfrak{c}(\underline{\underline{X}})_{i,j} \right>$
where
\begin{equation}
\mathfrak{c}(\underline{\underline{X}})_{i,j} :=  
\frac{1}{T} \sum_{t=1}^T X_{i,t} X_{j,t}
\label{eq:DefI}
\end{equation}
and the average is over the random ensemble $X_{i,t}$. We have:
\begin{eqnarray}
\rho_\Lambda(\lambda) &= &
\mbox{lim}_{\epsilon \rightarrow 0} \sum_{i=1}^N \delta_\epsilon(\lambda - \lambda_i)
= \mbox{lim}_{\epsilon \rightarrow 0}
\sum_{i=1}^N \mbox{Im$\left[G(\lambda - \lambda_i + \imath \epsilon)\right]$}\\
&=& \mbox{lim}_{\epsilon \rightarrow 0}
\mbox{Tr}
\left[
\mbox{Im}
\left[
G(\lambda 1 - \underline{\underline{\mathfrak{c}}} + \imath \epsilon)
\right]\right]
\end{eqnarray}

\begin{lemma}
The resolvent function $G(z)$ 
has
$\mbox{res}_{z = 1} G(z) = 1/2\pi$.
Here $\mbox{res}$ denotes a residue.
\end{lemma}

\begin{proof}
The function $G(z)$ can be expanded in a Laurent series around $z = 0$. 
This means that
\begin{equation}
\exists_{k \ge 0}\quad G(z) = \sum_{j=1}^k \frac{a_j}{z^j} + \mathcal{G}(z)
\label{eq:ResolventLaurExp}
\end{equation}
where the function $\mathcal{G}(z)$ is analytic.

We need to prove that $\mbox{Im}\left[\int\limits_{-\infty}^\infty G(x + \imath \epsilon)\right] = 1$
for $\epsilon >0$.
We consider at first the non-analytic term in (\ref{eq:ResolventLaurExp}).
\begin{equation}
\mbox{Im}\left[\int\limits_{-\infty}^\infty \frac{1}{(x + \imath \epsilon)^j} dx\right] = 
\left\{ \begin{array}{rr} 
         1 & \quad j=1 \\
	 0 & \quad j>1 
	 \end{array}
	 \right.
\label{eq:ImIntegral}
\end{equation}
where the above equality in  (\ref{eq:ImIntegral}) is straightforward
and the lower equality is derived by means of the Cauchy integral theorem. 
Now we analyse the analytic term. For $R >0$ we compute
\begin{equation}
\int\limits_{-R}^R \mathcal{G}(z + \imath \epsilon) dz = 
\int\limits_{-R + \imath\epsilon}^{R + \imath\epsilon} \mathcal{G}(z) dz = 
\int\limits_{-R}^R \mathcal{G}(z) dz  + 
\int\limits_{0}^\epsilon \left[ \mathcal{G}(R + \imath \epsilon) - \mathcal{G}(-R + \imath \epsilon) \right] d\xi 
\label{eq:ContourInt}
\end{equation}
where the last equality in (\ref{eq:ContourInt}) follows from the application
of the Cauchy integral theorem to a contour consisting of four
intervals $[-R,R]$, $[R,R+\imath\epsilon]$, $[R+\imath\epsilon,-R+\imath\epsilon]$
and $[-R+\imath\epsilon,-R]$. 
From the last expression on the right hand side in (\ref{eq:ContourInt})
we see that the imaginary part of that integral is $O(\epsilon)$
and hence disappears when $\epsilon \rightarrow 0$.
This fact and (\ref{eq:ImIntegral}) suffices to finish the proof.
\end{proof}

The class of functions $G(z)$ becomes narrowed down subject to the following
condition:
\begin{lemma}
The resolvent function $G(z) = 1/z$,
if and only if
\begin{equation}
\forall_{n \in \mathbb{N}} \frac{G^{(n)}(z)}{n!}(-1)^n = ( G(z) )^{n+1}
\label{eq:ResForm}
\end{equation}
\end{lemma}
The necessity follows in a straightforward manner from substituting
$G(z) = 1/z$ into (\ref{eq:ResForm}).
To prove the sufficiency we take $x \in \mathbb{R}$, multiply both sides
of (\ref{eq:ResForm}) by $x^n$ and sum over $n=0,1,\dots,\infty$.
Since the left hand side is the Taylor expansion of $G(z - x)$ around $z$
and the right hand side form a geometric series we obtain a functional equation:
\begin{equation}
G(z - x) = \frac{G(z)}{1 - x G(z)} = \frac{1}{\frac{1}{G(z)} - x}
\label{eq:ResEqI}
\end{equation}
We substitute $x = (1-\alpha) z$ into (\ref{eq:ResEqI})
for some $\alpha \in \mathbb{R}_+$ and get:
\begin{eqnarray}
\frac{1}{G(\alpha z)} &=& \frac{1}{G(z)}  + (\alpha - 1) z \label{eq:ResEqIIa} \\
&=& (\alpha - 1)\left( z + \frac{z}{\alpha} + \frac{z}{\alpha^2} + \dots\right) 
+ \frac{1}{G(0)} \label{eq:ResEqIIb} \\
&=& (\alpha - 1) z \frac{1}{1 - \frac{1}{\alpha}} = \alpha z \label{eq:ResEqIIc}
\end{eqnarray}
where in (\ref{eq:ResEqIIb}) we have iterated the equation (\ref{eq:ResEqIIa})
and we used the fact that $1/G(0)=0$. Finally in (\ref{eq:ResEqIIc}) we summed
a geometric series and completed the proof.

\section{Computation of the expansion of the resolvent\label{sec:Comput}}
We calculate a function 
\begin{equation}
\underline{\underline{g}}(z) = \left< G\left( z \cdot 1 - \frac{1}{T} \underline{\underline{X}}\cdot \underline{\underline{X}}^T \right)\right>
\label{eq:DefResolvent}
\end{equation}
where the average is performed over random variables
$(\underline{\underline{X}})_{i,\theta} \sim q-\mbox{Exp}$ that are correlated in $i$
and in time $\theta$ and $G(z)$ is a resolvent function.
The function $\underline{\underline{g}}(z)$ is termed the resolvent.
For the purpose of the calculation we fix $\vec{\xi}_{i,t}\in \mathbb{R}^{D}$, 
we project the joint pdf
$\rho_{\vec{X}}(\vec{x})$ onto Gaussians $\mbox{Normal(0,$\sigma_D^2/\xi_{i,t}^2$)}$
using the integral representation (\ref{eq:Projection}),
average over the Gaussians and obtain
\begin{equation}
\left<X_{i,\theta} X^T_{\theta',j}\right> = 
C_{i,\theta}^{j,\theta'}(\vec{\xi}) = \mathop{\sum_{p=1,\dots,N}}_{\lambda=1,\dots,T} 
                                 O_{i,\theta}^{p,\lambda} \frac{\sigma_D^2}{\xi_{p,\lambda}^2}
				 O_{p,\lambda}^{j,\theta'} 
\label{eq:CorrelDef}
\end{equation}
Note that the two-point correlation function (\ref{eq:CorrelDef}) does not 
factorize into functions depending on times $\theta,\theta'$ and on $i,j$ only.
It factorises for $|\xi_{i,t}| = 1$ if the rotation tensor $O_{i,t}^{p,\lambda}$
factorises, ie $O_{i,t}^{p,\lambda} = I_i^p T_t^\lambda$.
For generic value of $\vec{\xi}$, however, correlations in $i$ and in time are coupled with each other. 
In the following we sum over repeated indices (use Einstein's summation convention). 
The N-type indices and the T-type indices (running from one to N and to T)
are denoted by Latin and by Greek letters respectively.
We expand the resolvent function $G()$ in a series around $z$. We get:
\begin{eqnarray}
\lefteqn{
g(z)_{i,j} = \sum_{n=0}^\infty \frac{G^{(n)}(z)}{n!} (-1)^n 
             \left<
  \left(\underline{\underline{X}} \cdot \underline{\underline{X}}^T\right)^n_{i,j}
             \right>} \label{eq:ExpansionI} \\
&&= \sum_{n=0}^\infty \frac{G^{(n)}(z)}{T^n n!} (-1)^n \delta_{i,i(1)}
             \left<
  \left(\underline{\underline{X}} \cdot \underline{\underline{X}}^T\right)^n_{i(1),i(2 n+1)}
             \right> \delta_{i(2 n+1),j} \label{eq:ExpansionII} \\
&&=  \sum_{n=0}^\infty \frac{G^{(n)}(z)}{T^n n!} (-1)^n \delta_{i,i(1)}
             \left<
  \prod_{j=1}^{n} \left( X_{i(2 j-1),\xi(2 j-1)} X^T_{\xi(2 j-1),i(2j+1)} \right)
             \right> \delta_{i(2 n+1),j} \label{eq:ExpansionIII} \\
&&=  \sum_{n=0}^\infty \frac{G^{(n)}(z)}{T^n n!} (-1)^n \delta_{i,i(1)}
             \left<
   \prod_{j=1}^{n} \left( X_{i(2 j-1),\xi(2 j-1)} \delta_{\xi(2 j-1),\xi(2 j)} 
                            X^T_{\xi(2 j),i(2j)} 
                            \delta_{i(2j),i(2j+1)}
		    \right)
             \right>
	     \delta_{i(2 n+1),j} 
             \label{eq:ExpansionIV} \\
&&= \sum_{n=0}^\infty 
    \sum_{\sigma}
             \frac{1}{z} \delta_{i,i(1)}
             \left<
	     \left(
\prod_{j=1}^{n} 
C_{i(\mathfrak{c}^{(j)}_1),\xi(\mathfrak{c}^{(j)}_1)}
^{i(\mathfrak{c}^{(j)}_2),\xi(\mathfrak{c}^{(j)}_2)}
             \right)
	   \prod_{j=1}^{n}\left(\frac{1}{T}\delta_{\xi(2 j-1),\xi(2 j)}\right)
           \prod_{j=1}^{n-1}\left(\frac{1}{z}\delta_{i(2j),i(2j+1)}\right)
		    \right>
		    \frac{1}{z} \delta_{i(2 n),j}
              \label{eq:ExpansionV}  
\end{eqnarray}
In (\ref{eq:ExpansionII}) we introduced two new N-type indices
$i(1)$ and $i(2 n+1)$; in (\ref{eq:ExpansionIII}) we expanded the $n$-th power 
$\left(\underline{\underline{X}} \cdot \underline{\underline{X}}^T\right)^n_{i(1),i(2 n+1)}$
and we introduced
indices $i(1),i(3),\dots,i(2 n-1)$ together with $\xi(1),\dots,\xi(2 n-1)$.
In (\ref{eq:ExpansionIV}) we inserted $n$ T-type Kronecker delta functions between the
ordered $X,X^T$ pairs and $n-1$ N-type delta functions between the pairs $X^T,X$. 
This resulted in introducing indices $\xi(2),\dots,\xi(2n)$ and indices
$i(2),\dots,i(2n)$.

Finally in (\ref{eq:ExpansionV}) we made use of the Wick theorem (\ref{eq:ManyPointCorrel}) 
(the sum over $\sigma$ runs over all $2n$-permutations composed 
entirely of cycles $\mathfrak{c}^{(j)}$ of length two), 
of Lemmas 3.1 and 3.2 and we eliminated the index $i(2n+1)$.
Note that the sum over $\sigma$ contains $(2 n - 1)!!$ terms each containing
$4 n$ variable  indices $i(1),\dots,i(2n)$ and $\xi(1),\dots,\xi(2n)$
and two fixed indices $i$ and $j$.

\begin{figure}[tbh]
\centerline{\psfig{figure=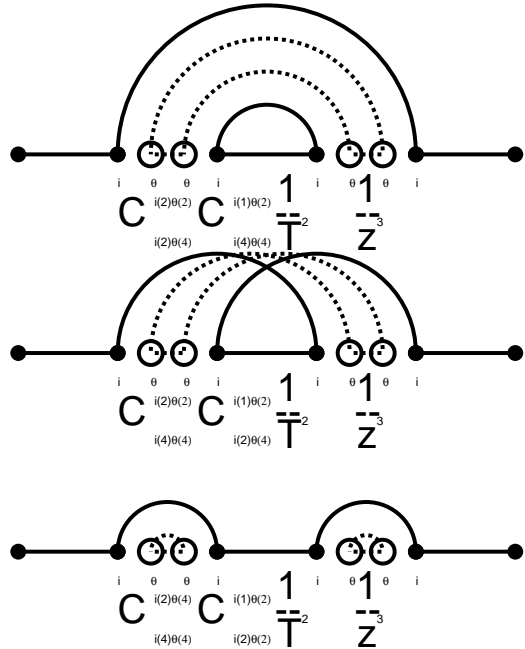,width=0.5\textwidth}}
\caption{All diagrams of the second order that contribute to the expansion of the resolvent and their weights.
\label{Fig:GraphsII}}
\end{figure}

\begin{figure}[tbh]
\centerline{\psfig{figure=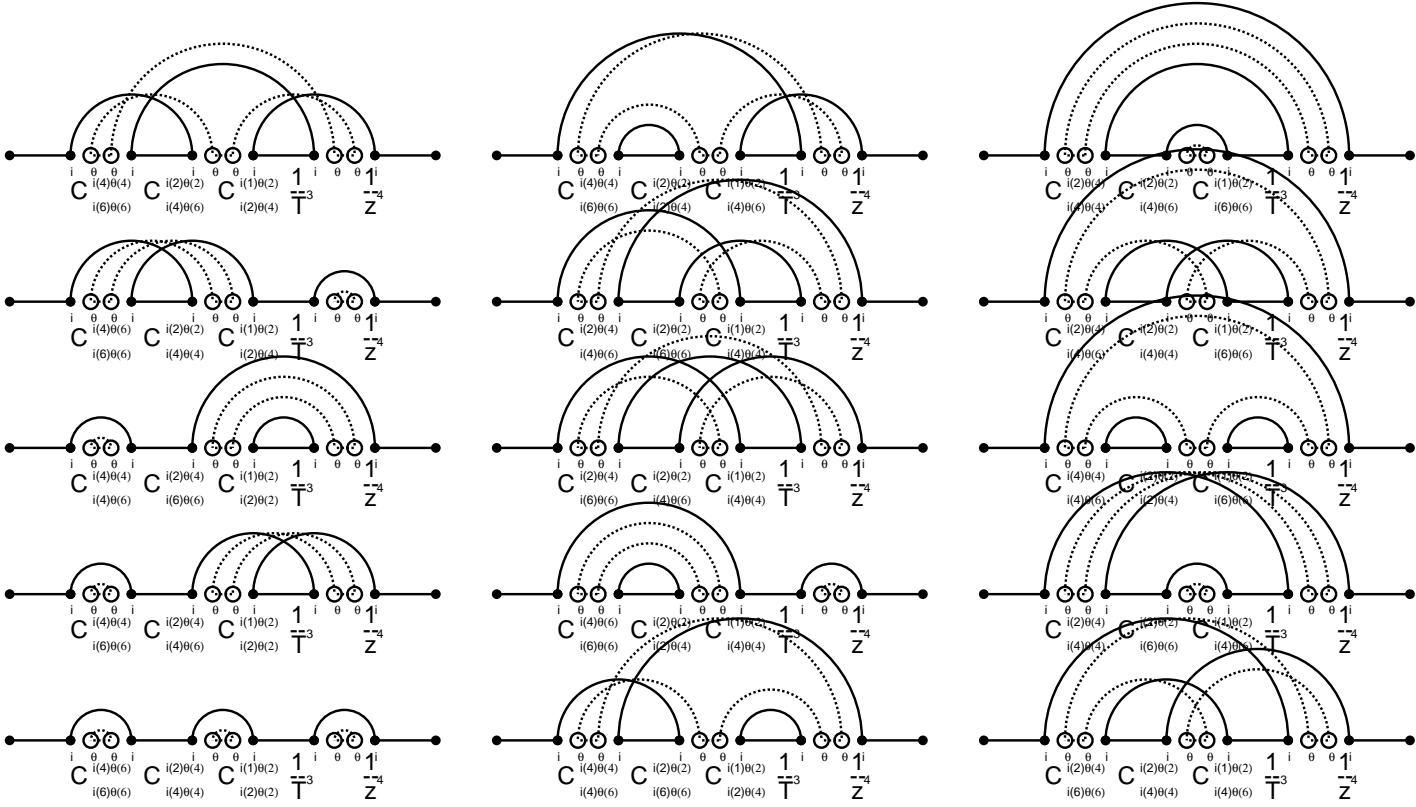,width=1.0\textheight,angle=-90 }}
\caption{The same as in Fig. \ref{Fig:GraphsII} but for diagrams of third order.
\label{Fig:GraphsIII}}
\end{figure}

Now we construct a pictorial representation (Feynmann diagrams)
\cite{Hooft,KleinertPhi4Theories,BurdaGoerlich,SenguptaMitra} of terms in the sum
(\ref{eq:ExpansionV}) according to a following recipe. 
The $i(\dots)$ indices are denoted by bullets $\bullet$,
and the $\xi(\dots)$ indices are denoted by open circles $\circ$.
The factors in the product on the right hand side in (\ref{eq:ExpansionV})
are assigned to graphs as follows:

\begin{center}
\begin{tabular}{cc}
Factor                   & Graph \\
$C_{i(1),\theta(1)}
^{i(2),\theta(2)}$          & \hbox{\psfig{figure=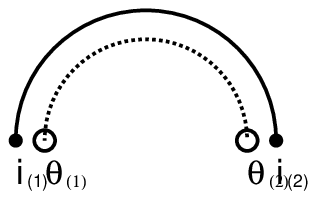,width=0.2\textwidth}} \\
$\frac{1}{z} \delta_{\theta(1),\theta(2)}$ & \hbox{\psfig{figure=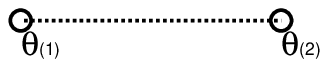,width=0.2\textwidth}} \\
$\frac{1}{N} \delta_{i(1),i(2)}$ & \hbox{\psfig{figure=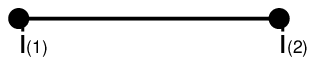,width=0.2\textwidth}} 
\label{eq:BuildBlocks}
\end{tabular}
\end{center}
In this way the sum (\ref{eq:ExpansionV}) can be represented as a sum over graphs
that are constructed from building blocks (\ref{eq:BuildBlocks}) 
in such a way that for every node, except for the nodes $i$ and $j$ (external nodes),
there are exactly two edges abutting at it (this follows from the Einstein's
summation convention).
All graphs contributing to the second and to the third order of the expansion
are listed, together with their weights, in Figs. \ref{Fig:GraphsII} and 
\ref{Fig:GraphsIII}.

Each graph consists of a number of closed solid, a number of closed dashed loops 
and of a solid line that starts at index $i$ and ends at $j$. 
Since a closed loop corresponds to a contraction 
(setting two indices of the tensor equal and summing over them) 
$\mathcal{A}_{i(1),\theta}^{i(2),\theta}$ of a tensor 
$\mathcal{A}_{i(1),\theta(1)}^{i(2),\theta(2)}$ 
that is constructed by multiplying the weights of the edges of the graph,
the weight of the closed loop is proportional to $N$ and to $T$
for solid and a dashed closed loops respectively. 
In other words the weight of a closed solid (dashed) loop
is equal to a trace of a $N\times N$ or ($T\times T$) matrix
and thus proportional to $N$ or ($T$).
In the following we assume that $N/T = r$ is fixed
and investigate the expansion (\ref{eq:ExpansionV}) 
in the limit $N\rightarrow \infty$.
Only such graphs contribute to the expansion whose number of loops 
(either dashed or solid) is equal to the order of the expansion $n$ (planar graphs).
Graphs that consist of intersecting lines, like the second graph from the top
in Fig.\ref{Fig:GraphsII} are negligible in the limit $N \rightarrow \infty$
since their weight is inverse proportional to a certain power of $N$.

\begin{figure}[tbh]
\centerline{\psfig{figure=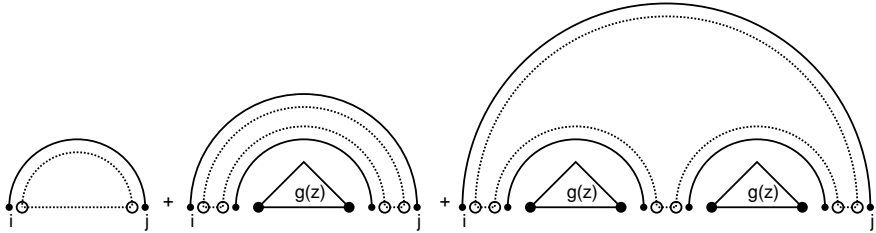,width=0.95\textwidth}}
\caption{Calculation of the self energy as a geometric sum of the resolvent contributions.
\label{Fig:SelfEnergyCalc}}
\end{figure}

We define one-line irreducible graphs as  
graphs that cannot be split into two distinct graphs by cutting
a certain edge. The usefulness of this definition follows from the fact
that the weight of a graph 
(the term in the sum (\ref{eq:ExpansionV}) for fixed $n$ and fixed $\sigma$) 
factorizes
into a product of weights corresponding to one-line irreducible components.
Note that this would not be the case if the weight depended explicitly
on $n$ (e.g. the function $G(z)$ satisfied Lemma 3.1 but not Lemma 3.2)
or if it depended on some function of indices $i(\dots)$ or $\xi(\dots)$).
Denoting by $\underline{\underline{\Sigma}} = \left\{\Sigma_i^j \right\}_{i,j}$ 
the sum of weights of all one-line irreducible
graphs (self-energy) 
with external nodes $i$ and $j$ we realize from (\ref{eq:ExpansionV})
and from Fig.\ref{Fig:SelfEnergy} that the resolvent is a sum of a geometric
series in self-energy:
\begin{figure}[tbh]
\centerline{\psfig{figure=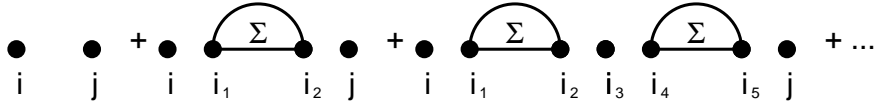,width=0.95\textwidth}}
\caption{Computation of the resolvent as a geometric sum of self-energy graphs.
\label{Fig:SelfEnergy}}
\end{figure}
\begin{equation}
\underline{\underline{g}}(z) = 
\frac{1}{z} 1 + \frac{1}{z^2} \underline{\underline{\Sigma}} + \frac{1}{z^3} \underline{\underline{\Sigma}}^2 + \dots =
\left( z 1 - \underline{\underline{\Sigma}} \right)^{-1}
\label{eq:SelfEnergy}
\end{equation}
This result is quite general, 
ie it holds also if non-planar graphs are taken into account.
However, the self-energy can only be calculated in the case when non-planar
graphs are neglected as we see in Fig.\ref{Fig:SelfEnergyCalc} and in equation (\ref{eq:SelfEnergyCalc}).
\begin{eqnarray}
&&\Sigma_i^j = \frac{1}{T} C_{i,\theta}^{j,\theta} +
\frac{1}{T^2} C_{i,\theta_1}^{j,\theta_2} C_{p,\theta_1}^{q,\theta_2} g(z)_{p,q} +
\frac{1}{T^3} C_{i,\theta_1}^{j,\theta_3} C_{p_1,\theta_1}^{p_2,\theta_2} g(z)_{p_1,p_2} C_{p_3,\theta_2}^{p_4,\theta_3} g(z)_{p_3,p_4}  + \dots \nonumber \\ 
&&= C_{i,\theta_1}^{j,\theta_2} 
\left((T 1 - \underline{\underline{\mathcal{B}}})^{-1}\right)_{\theta_1}^{\theta_2}
\label{eq:SelfEnergyCalc}
\end{eqnarray}
where $\mathcal{B}_{\theta_1}^{\theta_2} :=  C_{p,\theta_1}^{q,\theta_2} g(z)_{p,q}$.
Denoting $i=i(0)$ and $j=i(n)$ we write the result in a compact way:
\begin{equation}
g(z)_{i,j} = \sum_{n=0}^\infty \frac{1}{z^{n+1}} 
             \left(\prod_{q=0}^{n-1} \Sigma_{i(q)}^{i(q+1)}\right)
\label{eq:SelfEnergyResultI}	     
\end{equation}
\begin{equation}
\Sigma_i^j = \sum_{m=0}^\infty 
             \frac{C_{i,\theta(1)}^{j,\theta(m+1)}(\vec{\xi})}{T^{m+1}} 
	     \prod_{q=1}^m C_{j(2q - 1),\theta(q)}^{j(2q),\theta(q+1)}(\vec{\xi}) 
	                   g(z)_{j(2q-1),j(2q)}
\label{eq:SelfEnergyResultII}
\end{equation}
and we make following comments:
\begin{enumerate}
\item The results (\ref{eq:SelfEnergyResultI}) and (\ref{eq:SelfEnergyResultII})
      are to be understood as follows. We fix 
      $\vec{\xi} = \left(\xi_{1,1},\dots,\xi_{N,T}\right) \in \mathbb{R}^{N T \cdot D}$
      and for a given rotation tensor $\underline{\underline{O}}$ and $\sigma_D$
      we compute the correlation tensor $\underline{\underline{C}}(\vec{\xi})$ from
      (\ref{eq:DefinitionCorrelVar}), we insert the self-energy
      from equation (\ref{eq:SelfEnergyResultII}) into (\ref{eq:SelfEnergyResultI})
      and we iterate the result until convergence, 
      in the expansion in $1/z$ to a given order,  
      is obtained. Then we average the result over $\vec{\xi} \in \mathbb{R}^{N T \cdot D}$
      according to the rule (\ref{eq:Weight}). Since the weight in (\ref{eq:Weight})
      factorises and the resolvent as a function of $\vec{\xi}$ 
      is a polynomial of inverse powers of $\xi_{i,t}$ the weighting can be done analytically.
\item The correlation tensor $\underline{\underline{C}}$ is not a tensor product
      of $i$- and time $t$- dependent matrices (does not factorize)
      (as noted in the paragraph under equation (\ref{eq:CorrelDef}))
      even if the rotation tensor $\underline{\underline{O}}$ factorises.
\item The result (\ref{eq:SelfEnergyResultII}) is only valid in the limit $N\rightarrow \infty$. For finite values of $N$ there will be corrections proportional to
inverse powers of $N$, corrections resulting from non-planar graphs (see Figs.\ref{Fig:GraphsII} and \ref{Fig:GraphsIII}). 
\end{enumerate}

\section{The direct and inverse problems and the moment expansion of the resolvent}
The purpose of this section is to make connection between quantities that are measured
from financial time series, namely estimators of correlations 
$\mathfrak{c} := \mathfrak{c}(\underline{\underline{X}})_{i,j}$ 
(definition (\ref{eq:DefI})),
and between the underlying rotation tensor $O_{i,\theta}^{p,\lambda}$ and
the variance $\sigma_D$ of the q-Exponential distribution.
We define spectral moments $m_n$ of the estimator of correlations
$\mathfrak{c}$ as traces of powers of the estimator:
\begin{equation}
m_n := \frac{1}{N} \mbox{Tr$\left[\mathfrak{c}^{n}\right]$}
\end{equation}
for $n=0,1,2\dots$.
Having done that we readily see from the definition of the resolvent 
(\ref{eq:DefResolvent}) 
that the trace of the resolvent 
$t(z) := \frac{1}{N}\mbox{Tr$\left[ \underline{\underline{g}}(z) \right]$}$
is a generating function of the spectral moments 
$t(z) = \sum_{n=0}^\infty m_n/z^{n+1}$

Following the terminology from \cite{JurkiewiczRes} we define two problems
the direct and the indirect one.
The direct problem consists in  computing the spectral moments
from the rotation tensor $O_{i,\theta}^{p,\lambda}$ and the variance $\sigma_D$.
The indirect problem, which is more interesting from the point of view 
of applications in quantitative finance, is defined as determining the 
correlation tensor and the variance from the sole knowledge of the spectral moments.
To what extent it is possible, 
what additional assumptions about the structure of the tensor have to be made before 
deriving a numerical algorithm and 
what is the error estimate in the algorithm will be discussed in future work.

\subsection{Finding the resolvent}
We solve equations (\ref{eq:SelfEnergyResultI}) and (\ref{eq:SelfEnergyResultII})
for $g(z)_{i,j}$
according to the recipe in point (1) at the end of section \ref{sec:Comput}
and obtain following results:
\begin{tabular}{cc}
Order $n$ & The resolvent $g^{(n)}(z)_{i,j}$ \\ \hline 
0         & \begin{minipage}{0.8\textwidth}
             \begin{equation}\delta_{i,j}\frac{1}{z}\end{equation}        
	    \end{minipage}                 \\ \hline
1         & \begin{minipage}{0.8\textwidth}
             \begin{equation}\frac{1}{z}\delta_{i,j} + 
	     \frac{1}{z^2}\left( \frac{1}{T} C_{i,\theta_1}^{j,\theta_1} \right)\end{equation} 
	    \end{minipage}                 \\ \hline
2         & \begin{minipage}{0.8\textwidth}
             \begin{equation}\frac{1}{z}\delta_{i,j} + 
	     \frac{1}{z^2}\frac{1}{T} \left( C_{i,\theta_1}^{j,\theta_1} \right) +
             \frac{1}{z^3}\frac{1}{T^2}\left( 
               C_{i,\theta_1}^{j,\theta_2} C_{j_1,\theta_1}^{j_1,\theta_2} +
	       C_{i,\theta_1}^{k,\theta_1} C_{k,\theta_2}^{j,\theta_2} 
	      \right)
	     \end{equation}
	    \end{minipage}                 \\ \hline
3         & \begin{minipage}{0.8\textwidth}
             \begin{eqnarray}
	     \lefteqn{
              \frac{1}{z}\delta_{i,j} + 
	     \frac{1}{z^2}\frac{1}{T} \left( C_{i,\theta_1}^{j,\theta_1} \right) +
             \frac{1}{z^3}\frac{1}{T^2}\left( 
               C_{i,\theta_1}^{j,\theta_2} C_{j_1,\theta_1}^{j_1,\theta_2} +
	       C_{i,\theta_1}^{k,\theta_1} C_{k,\theta_2}^{j,\theta_2} 
	      \right)} \\
	      &&+\frac{1}{z^4}\frac{1}{T^3}
	      (
              C_{i,\theta_1}^{j,\theta_2} C_{j_1,\theta_1}^{j_2,\theta_2} C_{j_1,\theta_3}^{j_2,\theta_3}  +
C_{i,\theta_1}^{j,\theta_3} C_{j_1,\theta_1}^{j_1,\theta_2} C_{j_3,\theta_2}^{j_3,\theta_3} +
C_{i,\theta_1}^{k,\theta_1} C_{k,\theta_3}^{j,\theta_4} C_{j_1,\theta_3}^{j_1,\theta_4} + \nonumber \\
&&
C_{i,\theta_3}^{k,\theta_4} C_{k,\theta_1}^{j,\theta_1} C_{j_1,\theta_3}^{j_1,\theta_4} +
C_{i,\theta_1}^{k_1,\theta_1} C_{k_1,\theta_2}^{k_2,\theta_2} C_{k_2,\theta_3}^{j,\theta_3}
	      ) \label{eq:eq:ResolventOrder3}
	      \end{eqnarray}
	      \end{minipage}
\end{tabular}
Comparing the coefficients of the expansion  (\ref{eq:eq:ResolventOrder3})
in powers of $1/z$ with the spectral moments $m_n$ of the estimator of the correlations
we obtain a set of non-linear equations that relate certain contractions of the 
correlation tensor to the spectral moments.
If we assumed that the correlation tensor factorized, which is not the case as we discussed 
in point (1) in section \ref{sec:Comput}, then we would have obtained equations (34)
from \cite{JurkiewiczRes}, equations that relate the spectral moments to moments
of the underlying correlation matrices both in $i$ and in time.
In our case the relations are averaged over $\vec{\xi}$ and will be related to some
contractions of the rotation tensor $\underline{\underline{O}}$.
Before proceeding further we note that the relations solve the direct problem
but they do not provide enough information to solve the indirect problem.

\section{Conclusions} We have derived a variant of the Wick theorem
that expresses the many-point correlation function of $q$-Exponentialy distributed 
random variables through two-point correlation functions.
This theorem will be used for solving the indirect problem
in quantitative finance, ie for determining the 
correlations of time series from the knowledge of the spectral moments of the estimator of covariance.

\section{Acknowledgments} We thank Hagen Kleinert 
www.physik.fu-berlin.de/$\sim$kleinert for suggesting this problem and for discussions.


\end{document}